\documentclass[seceq]{ptptex}

\usepackage{graphicx}

\title{
Primordial Gravitational Waves from Inflation and Preheating%
}

\author{
Juan \textsc{Garc\'ia-Bellido}%
}

\inst{
Departamento e Instituto de F\'isica Te\'orica UAM-CSIC, \\
Universidad Aut\'onoma de Madrid, Cantoblanco 28049 Madrid, Spain\\
D\'epartement de Physique Th\'eorique, Universit\'e de Gen\`eve, \\
24 quai Ernest Ansermet, CH--1211 Gen\`eve 4, Switzerland
}

\abst{
There are at least three cosmic backgrounds of primordial gravitational waves
coming from inflation: those produced during inflation and associated with the 
stretching of quantum modes; those produced at the violent stage of preheating
after inflation; and those associated with the self-ordering of Goldstone modes if inflation
ends via a global symmetry breaking scenario, like in hybrid inflation. Each background 
has its own characteristic spectrum with specific features. We describe in detail the
origin of extra peaks in the PGW background from preheating in the case in which
inflation ends with the formation of cosmic strings due to an Abelian Higgs model after
hybrid inflation. We then discuss the prospects for detecting each gravitational wave 
background, and distinguishing between them with a very sensitive probe, the
local B-mode of the cosmic microwave background polarization.
}

\begin{document}

\maketitle

\section{Introduction}

Cosmological Inflation~\cite{LindeBook,MukhanovBook} naturally generates a spectrum of density fluctuations
responsible for large scale structure formation which is consistent with the observed CMB 
anisotropies.\cite{Komatsu2010} It also generates a spectrum of gravitational waves, 
whose amplitude is directly related to the energy
scale during inflation and which induces a distinct B-mode polarization pattern in the 
CMB.\cite{DurrerBook} Moreover, Inflation typically ends in a violent process at preheating,\cite{preheating} 
where large density waves collide at relativistic speeds generating a stochastic background 
of GW~\cite{GWpreh} with a non-thermal spectrum characterized by a prominent peak at GHz 
frequencies for GUT-scale models of inflation (or at mHz-kHz for low scale models of inflation), 
and an amplitude proportional to the square of the mass scale driving/ending inflation. Such a 
background could be detected with future GW observatories like Adv-LIGO~\cite{ligo}, 
LISA~\cite{lisa}, BBO~\cite{bbo}, DECIGO~\cite{decigo}, etc. In case the end of inflation involves the
presence of gauge fields associated with the breaking of some symmetry, like in most scenarios of 
hybrid inflation, then the PGW spectrum presents an extra peak associated with the mass scale of
the corresponding gauge fields~\cite{DFG2010}. This would constitute a very clear signature of the
physics of reheating, and its detection would open a new window into the early universe.

Furthermore, if inflation ended with a global O(N) phase transition, like in certain scenarios of
hybrid inflation, then there is also a GWB due to the continuous self-ordering of the Goldstone 
modes at the scale of the horizon,\cite{Krauss} which is also scale-invariant on subhorizon 
scales,\cite{FFDGB} with an amplitude proportional to the quartic power of the symmetry breaking 
scale, that could be detectable with laser interferometers as well as indirectly with the B-mode 
polarization of the CMB.\cite{GBDFFK}

\begin{figure}
\vspace{-0.5cm}
\centerline{
\includegraphics[width=13.0 cm,height=10 cm]{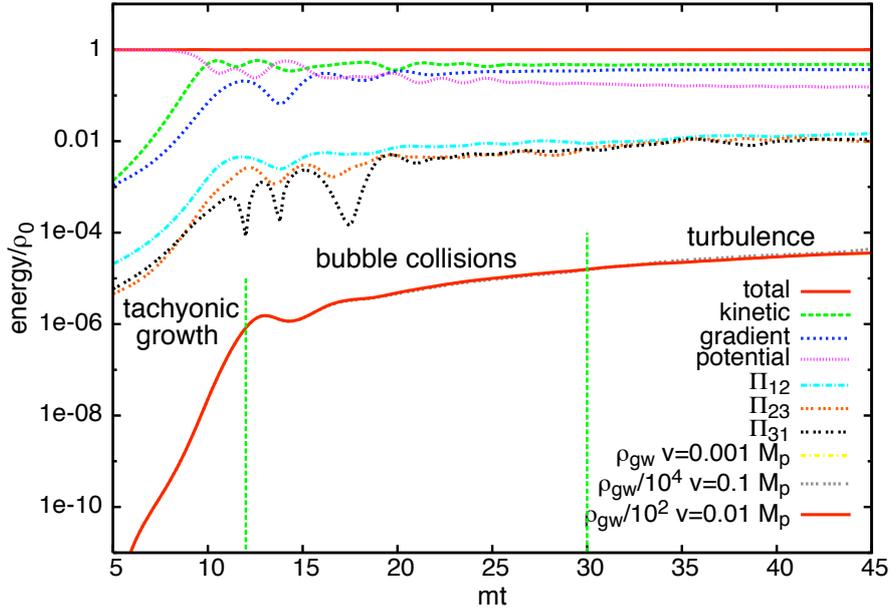}
}
\vspace{-0.5cm}
\caption{The time evolution of the GW energy density during the initial stages of preheating after
hybrid inflation, from Ref.[6]. Note the three stages of tachyonic growth, bubble collisions and turbulence.}
\label{fig:GWpreh}
\end{figure}

Gravitational waves produced during inflation arise exclusively due to the quasi-exponential expansion of the 
Universe~\cite{MukhanovBook}, and are not sourced by the inflaton fluctuations, to first order in 
perturbation theory. They have an approximately scale invariant and Gaussian spectrum whose 
amplitude is proportional to the energy density during inflation. GUT scale inflation has good chances 
to be discovered (or ruled out) by the next generation of CMB anisotropies probes, Planck~\cite{Planck}
COrE~\cite{Bpol} and CMBpol~\cite{CMBpol}.
At the end of inflation, reheating typically takes place in several stages. There is first a rapid (explosive)
conversion of energy from the inflaton condensate to the fields that couple to it. This epoch is known as 
preheating~\cite{preheating} and occurs in most models of inflation. It can be particularly violent in the
context of hybrid inflation, where the end of inflation is associated with a symmetry breaking scenario,
with a huge range of possibilities, from GUT scale physics down to the Electroweak scale. Gravitational 
waves are copiously produced at preheating from the violent collisions of high density waves moving
and colliding at relativistic speeds~\cite{GWpreh}, see Fig.~1. The GW spectrum is highly peaked at the 
mass scale corresponding to the symmetry breaking field, which could be very different from the Hubble scale.

In low-scale models of hybrid inflation it is possible to attain a significant GWB at the range of frequencies
and sensitivities of LIGO or BBO. The origin of these GW is very different from that of inflation. Here the
space-time is essentially static, but there are very large inhomogeneities in the symmetry breaking (Higgs) field
due to the random spinodal growth during preheating. Although the transition is not first order, ``bubbles" 
form due to the oscillations of the Higgs field around its minimum. The subsequent collisions of the 
quasi-bubble walls produce a rapid growth of the GW amplitude due to large field gradients, which 
source the anisotropic stress-tensor~\cite{GWpreh}. The relevant degrees of freedom are those of the 
Higgs field, for which there are exact analytical solutions in the spinodal growth stage, which later can be 
input into lattice numerical simulations in order to follow the highly non-linear and out-of-equilibirum stage 
of bubble collisions and turbulence. However, the process of GW production at preheating lasts only a 
short period of time around symmetry breaking. Soon the amplitude of GW saturates during the turbulent 
stage and then can be directly extrapolated to the present with the usual cosmic redshift scaling. Such a 
GWB spectrum from preheating would have a characteristic bump, worth searching for with GW observatories 
based on laser interferometry, although the scales would be too small for leaving any indirect signature in 
the CMB polarization anisotropies. Moreover, the mechanism generating GW at preheating is also active 
in models where the SB scenario is a local one, with gauge fields present in the plasma, and possibly 
related to the production of magnetic field flux tubes~\cite{PMFpreh}. In such a case, one could try to 
correlate the GWB amplitude and the magnitude and correlation length of the primordial magnetic field seed.

\section{Gravitational waves from Abelian Higgs Cosmic Strings}

Gravitational waves are a robust prediction of general relativity. There is indirect evidence of their existence
from inspiraling binary pulsars, although no single direct detection has been claimed. A stochastic background
of GW may soon be discovered, either directly with laser interferometer antennas or indirectly through the pattern
of polarization anisotropies they induce in the cosmic microwave background. Such a detection would open
a completely new and unexplored window into the Early Universe, possibly as rich as that which has been
recently revealed in the CMB. There are many sources of GW that can generate a stochastic backgrounds
and thus it is necessary to characterize those backgrounds with as much detail as possible. Apart from
known astrophysical point-like sources beyond the confusion limit (where we cannot resolve them), there
are also predictions for GWB from cosmic defects and hypothetical strongly first order phase transitions
in the Early Universe. Moreover, Cosmological Inflation makes a robust prediction of a stochastic GWB
produced during the quasi-exponential expansion of the Universe, with very specific spectral signatures:
a Gaussian, almost scale-invariant spectrum with an amplitude directly related to the energy scale of inflation.
If the scale is high enough (close to the Grand Unification Theories' scale) then these GW will also leave
an imprint in the (curl) polarization anisotropies of the CMB. Unfortunately, this GWB is still too weak to be
discovered with the near-future GW interferometric antennas, although it does cover a sufficiently broad
frequency range to be detectable by future GW Observatories (GWO) like BBO or DECIGO.

Furthermore, a robust prediction of inflation is that it must have ended, converting the vacuum energy
responsible for the tremendous expansion into the matter and radiation we observe today. Such a process,
known as reheating, is typically very violent and very inhomogeneous, with large density waves moving
at relativistic speeds and colliding among each other, thus converting a large fraction of their gradient energy
into gravitational waves. In some cases, the conversion is so sudden and violent that a significant fraction of
the total energy that goes into radiation ends in a stochastic background of GW, which could be detected in
the future. The energy spectrum of such a GWB is very non-thermal and far from scale invariant, but actually
peaked at a frequency which is related to the typical mass scale responsible for the end of inflation (either the
mass of the inflaton or that of the field that triggers the end of inflation, like in hybrid models), which could
be orders of magnitude smaller than the Hubble scale at the end of inflation. However, if the energy scale
of inflation is large (of order the GUT scale) then this stochastic GWB will be peaked at GHz frequencies,
far from the present sensitivity of GW interferometers. Nowadays, our only chance of detecting the GWB from
reheating is to consider the low-scale models of inflation - like hybrid models - with the appropriate parameters
to convert a large fraction of the initial vacuum energy into GW. The analyses done so far have
considered only scalar fields whose gradient energies source the anisotropic stresses needed for GW
production. However, vector fields (gauge or not gauge) are expected to be an even better source of GW, 
due to their anisotropic curl components, so that preheating scenarios with gauge fields may have a larger 
contribution to the GWB than scalar models. In fact, previous studies of gauge fields at preheating, 
in the context of Electroweak Baryogenesis and in the generation of Primordial Magnetic Fields~\cite{PMFpreh}, have identified
the formation of long-wave semiclassical gauge field configurations like sphalerons and helical magnetic
flux tubes which evolve with time very anisotropically and could contribute significantly to the production of GW.

In a recent paper~\cite{DFG2010}, 
we developed a formalism to calculate the production of GW by coupled systems of scalar 
and gauge fields on the lattice. The numerical method that we have constructed can be applied to different 
sources of GW where out-of-equilibrium gauge fields play an important role, such as thermal phase transitions, 
cosmological networks of local defects or non-perturbative decays of scalar condensates into gauge fields. 
We have studied in detail the dynamics and the production of GW during preheating after hybrid inflation, 
in the context of abelian-Higgs models that go through dynamical symmetry breaking triggered by the
expectation value of the inflaton field. As the inflaton is driven slowly (as opposed to a quench)
below the critical value, the mass squared of the abelian Higgs field becomes negative and drives the
spinodal growth of long-wave modes of the Higgs. Since the Higgs is charged, its rapid growth induces
a corresponding growth of gauge field configurations. At the end of inflation there are no temperature
fluctuations that can induce over-the-barrier transitions. However, long-wave quantum fluctuations become
semi-classical and act as a stochastic force that allow transitions over the false vacuum and thus induce
(locally) the generation of a topological winding number of the Higgs field. After symmetry breaking,
there is not enough energy to unwind the Higgs phase, leaving behind a Nielsen-Olesen string.
Such cosmic string configurations can be seen explicitly in our spatial distributions of both Higgs and
gauge fields~\cite{DFG2010}. 
They play a crucial role in the production of GW at preheating and we observe that the spatial 
distribution of GW is indeed concentrated around the strings. Those strings will eventually decay (we see that 
they become wider and disperse away their energy density in the form of small-scale structures of the fields,
although the winding phase around the core of the string remains, since it is topologically stable), which
eventually shuts-off the GW production.

\begin{figure}
\centerline{
\includegraphics[width=13 cm,height=9 cm]{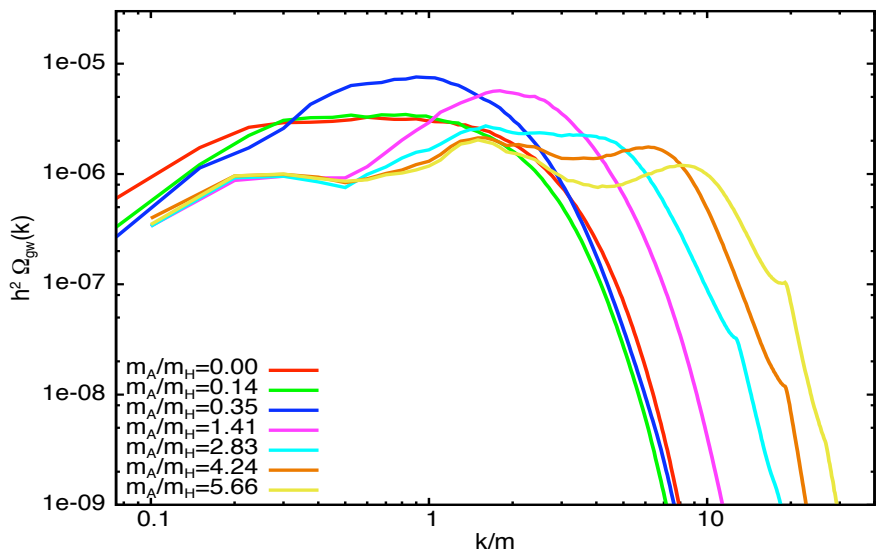}
}
\caption{The fraction of energy density in GW as a function of the typical momenta at preheating after inflation in the abelian-Higgs model for different ratios of the gauge field to Higgs mass. Note the existence of three peaks in the most extreme cases: an infrared peak due to straight segments of strings, and two more ultraviolet peaks, related to the Higgs and gauge field masses.}
\label{fig:spectraGW}
\end{figure}

The complicated dynamics occurring at preheating in this abelian Higgs-inflaton model has been studied
using both power spectra analyses, as well as the fields' distributions in configuration space, together with
histograms of the fields' values, as a function of time, in order to correlate the different features observed
and their evolution. The picture that arises is the following. At the end of inflation the tachyonic growth of
the Higgs Gaussian random field creates an inhomogeneous distribution of fields characterized by
``bubbles" of Higgs energy density that expand and collide. The gauge field concentrates at the valleys
between the bubbles, where the Higgs has low values, forming long flux tubes of magnetic energy density.
The dynamics of the bubles when they expand and collide leads to regions in space where the Higgs field 
reaches the false vacuum and there are over-the-barrier transitions, with topological windings associated 
with them. These Nielsen-Olesen vortices are connected with each other in a cosmic string which runs along 
the core of the magnetic flux tubes. There are strings that encompass the whole simulation box and even beyond, 
thanks to periodic boundary conditions. We observe this process both in configuration space and with the histograms 
of Higgs vevs.

We have followed the dynamics of the strings during and after the symmetry breaking, although still on time 
scales shorter than the Hubble time and on length scales smaller than the Hubble radius. Once the strings are 
formed, they evolve by increasing their size and shedding away layers of magnetic energy density. At the cores 
of the strings there always remain a thin magnetic flux line but the energy seems to pour away from the strings 
in the form of waves concentric with the string. Nevertheless, we observe (in a transverse plane to the string) 
that at the core of the string there remains a conserved winding number of the Higgs. We have followed this winding 
number up to long times and we confirm that it is still there, in spite of the fact that the magnetic flux tube is 
so dilute that we cannot see it coherently: it seems to have ``evaporated". What remains is a diffuse background of 
small-scale structures of the Higgs and gauge fields permeating the whole box, together with the remnants of the 
strings.

The formation, evolution and fragmentation of the strings are accompanied by a significant production of gravitational 
waves which inherit specific features from the string dynamics. In position space, we observe how the distribution of 
GW follows very closely the evolution of the strings, being first concentrated around the straight segments of strings, 
then fattening as the strings become wider and finally being dispersed over the lattice as the strings emit small-scale 
structures of the fields. In Fourier space, this dynamics is encoded into the successive appearance of 
very distinct peaks in the GW spectra. The position of each peak is directly related to the physical scales
in the problem: the Higgs mass, which governs the width and interactions of Higgs field's strings, the gauge field mass,
which governs the width and interactions of gauge field's strings, and the typical momentum amplified by tachyonic 
preheating, which determines the characteristic size of the bubbles when they collide and the correlation length of the 
straight segments of strings. The former two determine the peaks in the high momentum (UV) range of the spectrum, while 
the latter corresponds to the long-wave (IR) peak. The IR peak appears first, when the bubbles collide and the strings 
are formed, while the UV peaks are formed later on, when the strings evolve and decay into small-scale structures of the 
Higgs and gauge fields. When the different scales are close to each other, the different peaks are superimposed and 
the amplitude in GW increases. When the gauge coupling constant is significantly smaller than the Higgs' self-coupling, 
the results reduce to the GW spectra produced without gauge field, characterized by a single peak. 

We have calculated the GW spectra produced in this abelian Higgs model of preheating after hybrid inflation with state-of-the-art simulations, although still limited in spatial resolution and box sizes. In order to probe reliably the different scales in the problem in each simulation, we developed a lattice calculation of GW production with gauge fields that is accurate up to second order in the lattice spacing. The present-day frequency and amplitude of these GW are very sensitive to the model parameters and the frequency of the different peaks may differ by many orders of magnitude, as illustrated in Fig.~\ref{fig:spectraGW}. As in the same model with only scalar fields, very small coupling constants are still neccessary for these GW to fall into a frequency range that is accessible by interferometric experiments. Whether this is natural or not depends on the underlying theory for inflation and particle-physics models of hybrid inflation with such small coupling constants have indeed been already proposed in the literature, see~\cite{GWpreh} and references therein.  We also observed that the frequency of the IR peak in the GW spectrum can be smaller than the peak frequency produced in the same model with only scalar fields, so the gauge field may enlarge the regions of the parameter space that may lead to an observable signal. More generally, there are many other models of inflation and preheating where gauge fields may play an important role and wich may lead to GW that could be observed in the future.    

\begin{figure}
\centerline{
\includegraphics[width=13.0 cm,height=8 cm]{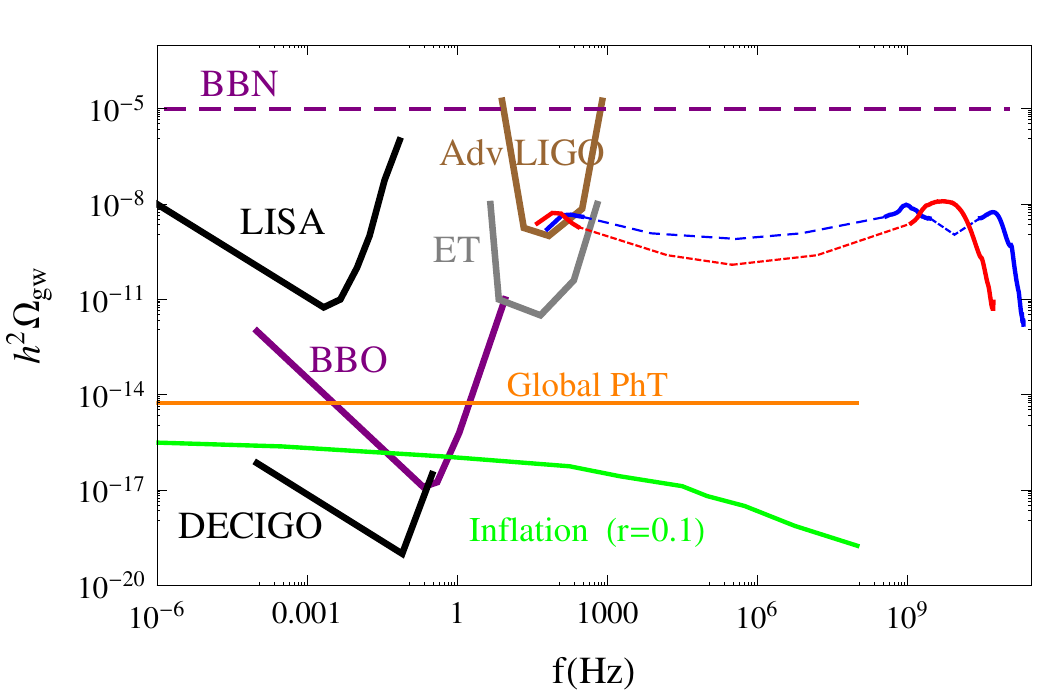}
}
\caption{The predicted stochastic background of GW from preheating in the abelian-Higgs model for two different sets of parameters, 
together with the expected sensitivity from future GWO like Advanced LIGO/VIRGO, LISA, ET, BBO and DECIGO. Note that in order to reach GWO sensitivity we had to extrapolate the position of the IR peaks and make an educated guess for the shape of the spectra between the peaks (dashed lines). Also plotted are the expected GWB from a global phase transition and that from an inflationary model with tensor to scalar ratio $r=0.1$.}
\label{fig:sensi}
\end{figure}

After preheating the system enters into a turbulent regime where, at least in the abelian Higgs model that 
we considered, gravitational waves are no longer produced and the GW energy density saturates. We expect this 
result to be rather generic in abelian scalar gauge theories, because in that case the gauge fields that are 
produced at preheating acquire a mass, either directly through the Higgs mechanism or due to their interactions 
with scalar fields' fluctuations. It would be interesting to study models with other gauge groups, like 
$SU(2) \times U(1)$, where gauge fields remain effectively massless after symmetry breaking. This may happen for instance during 
preheating after hybrid inflation close to the electroweak scale (possibly a secondary stage of inflation, 
not necessarily related to the CMB anisotropies, only responsible for reheating the universe),    
where the photon spectra may exhibit inverse cascade during the turbulent evolution towards thermal 
equilibrium~\cite{PMFpreh}. This could significantly lower the typical frequency of the resulting GW today and 
relax the conditions on the parameters for these GW to be observable. The details of the GW spectra produced from preheating should also be rather sensitive to the particular gauge group under consideration because this determines the nature of the defects that can be formed. The defects do not have to be stable since these GW are produced when they are being formed. 

An intriguing possibility is the following. Given that preheating is so extremely inhomogeneous, and since
these inhomogeneities get imprinted in the gravitational wave background, which immediately decouples
from the plasma, one may envision a stage of technological development in the not so far future in which
GWO with sufficient angular resolution may resolve the structures that gave rise to the GWB right at the
moment when the Universe reheated. In the usual preheating scenario at high energy scales with only
scalar fields, the physical structures will have today a size that is completely undetectable when projected
over the sky, and thus the GWB will look essentially homogeneous from Earth. However, if gauge fields with
long string-like configurations (of horizon size and possibly even with superhorizon correlations) were
behind the generation of the GWB, then one could expect to see inhomogeneities in the angular distribution
of those gravitational waves. In particular, an array of GWO could detect the string-like anisotropies in the
GWB across the sky. At the moment, the angular resolution of LIGO is not better than a degree projected
in the sky. However, in the future one could resolve much finer structures in the GWB thanks to a dense
network of ground-based laser interferometers~\cite{GWBanisotropies}. Thinking 
ahead of our times, it may not be unrealistic to imagine that in the not so far future the GWB will
be mastered with sufficient detail to resolve the anisotropies in this elusive background and thus recover
vital information about the physics responsible for the violent conversion of energy from inflation to a
radiation and matter dominated epoch (the Big Bang of the Old Theory). No other probe can give us so
much information, since GW decouple immediately upon production and thus retain the spatial and
energy distributions of the sources that produced them. We can compare with the Cosmic Microwave
Background, which gives us detailed information about the epoch of photon decoupling thanks to the
exquisite measurements of the angular correlation of both temperature and polarization anisotropies.
The CMB provides a snapshot of how the Universe was like 380,000 years after the Big Bang. On the other
hand, the GWB would open a window into the physics of the Big Bang itself, allowing us to infer from its
detailed features whether it was as violent and inhomogeneous as we predict, to determine what kind of
fields were present and whether the rich phenomenology that we associate with preheating (topological
defects, baryogenesis and/or leptogenesis, primordial magnetic seed creation, non-thermal production
of dark matter, etc.) was actually realized in nature.

\begin{figure}[ht]
\centerline{
\includegraphics[width=13.0 cm,height=9 cm]{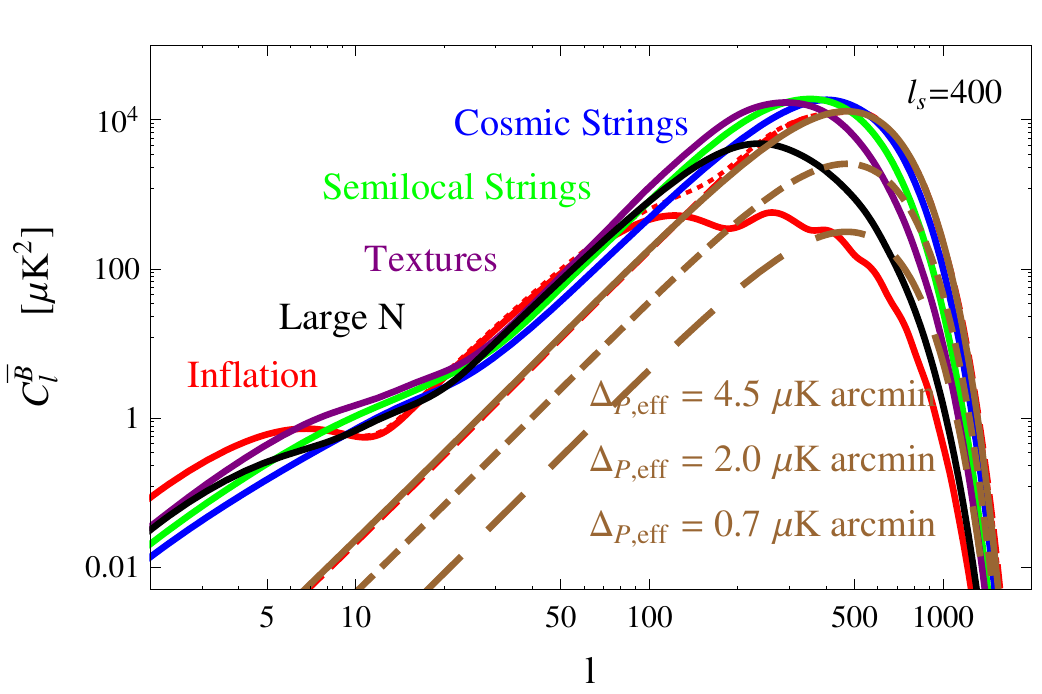}
}
\caption{The local $\tilde{B}$-polarization power spectra for tensor perturbations from inflation, 
cosmic strings, textures and the large-$N$ limit of thenon-linear sigma-model. All spectra are normalized 
such that they make up 10\% of the observed temperature anisotropy at $\ell=10$. Note that the lensed EE 
modes (red dashed line) contribute as colored noise to the local-BB angular power spectrum.
}
\label{fig:CBI}
\end{figure}

\section{Gravitational waves from global defects formed after inflation}

In the case that inflation ends with a global or local symmetry breaking mechanism, then there generically
appear cosmic defects associated with the topology of the vacuum manifold. For instance, global cosmic
strings are copiously produced during preheating if the Higgs field is a complex scalar with a U(1) global
symmetry~\cite{preheating}. In principle, all kinds of topological and non-topological defects could form at
the end of inflation and during preheating. Such defects will have contributions to all three different metric
perturbations: scalar, vector and tensor, with similar amplitudes~\cite{defrev}. In a recent work~\cite{FFDGB}, we analyzed 
the production of GW at preheating for a model with O(N) symmetry. The dynamics at subhorizon scales 
was identical to that of the usual tachyonic preheating. However, in this model even though the Higgs 
potential fixed the radial component to its vev, there remained the free (massless) Goldstone modes to
orient themselves in an uncorrelated way on scales larger than the causal horizon. In the subsequent 
evolution of the Universe, as the horizon grows, spatial gradients at the horizon will tend to reorder
these Goldstone modes in the field direction of the subhorizon domain. This self-ordering of the fields
induces an anisotropic stress-tensor which sources GW production. In the limit of large N components, 
it is possible to compute exactly the scaling solutions, and thus the amplitude and shape of the GWB
spectrum. It turns out that the GWB has a scale-invariant spectrum on subhorizon scales~\cite{Krauss} 
and a $k^3$ infrared tail on large scales~\cite{FFDGB}, which can be used to distinguish between inflation 
and these non-topological defects.

\begin{figure}[ht]
\centerline{
\includegraphics[width=13.0 cm,height=9 cm]{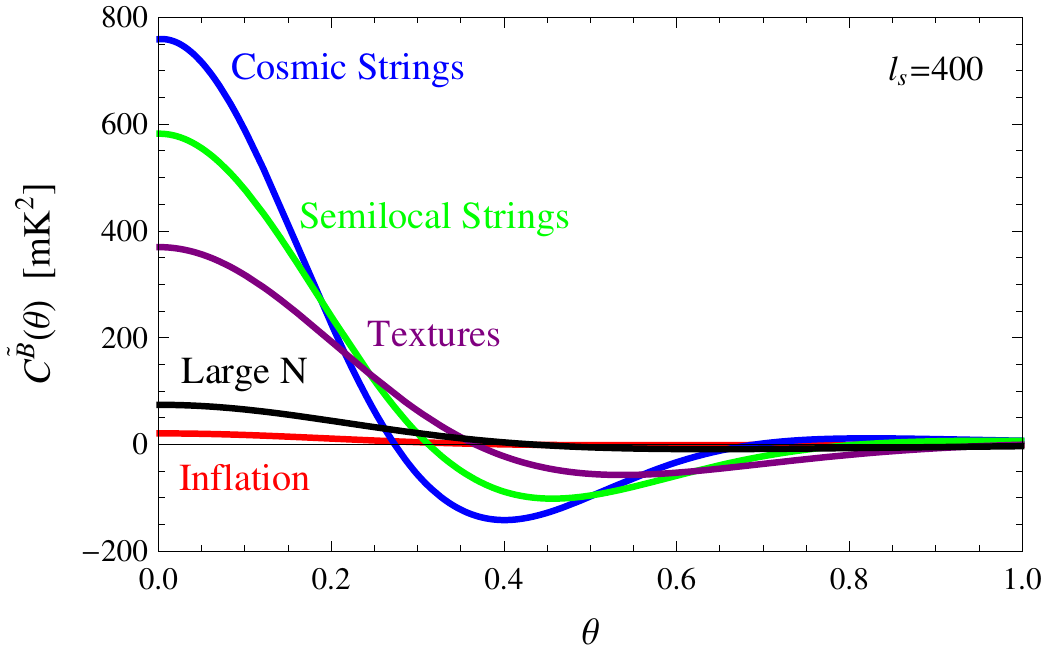}
}
\caption{The local $\tilde{B}$-polarization angular correlation functions for $\theta<1^o$ for inflation, 
cosmic strings, textures and the large-$N$ limit of NL sigma model.}
\label{fig:CB}
\end{figure}

Apart from the IR tail, the main difference between inflationary and global defect contributions to the
CMB anisotropies arises from the fact that defects generically contribute with all modes: scalar, vector 
and tensor modes, with similar amplitudes, while inflationary tensor modes could be negligible if the
scale of inflation is well below the GUT scale. Since (curl) B-modes of the polarization anisotropies
only get contributions from the vector and tensor modes, the detection of the B-mode from inflation
may be challenging, and dedicated experiments like Planck, COrE and CMBpol have been designed to look 
for them. On the other hand, defects' contribution to the temperature anisotropies have a characteristic
smooth hump in the angular power spectrum, which allows one to bound their amplitude (and thus the 
scale of symmetry breaking) below $10^{16}$ GeV.~\cite{defCMB} However, the contribution to the 
B-mode coming from defects have both tensor and vector components, and the latter can be up to ten 
times larger than the former, and actually peaks at a scale somewhat below the horizon at last scattering 
(in harmonic space the corresponding multipole is $\ell\sim1000$).

\begin{figure}
\centerline{
\includegraphics[width=13 cm,height=10 cm]{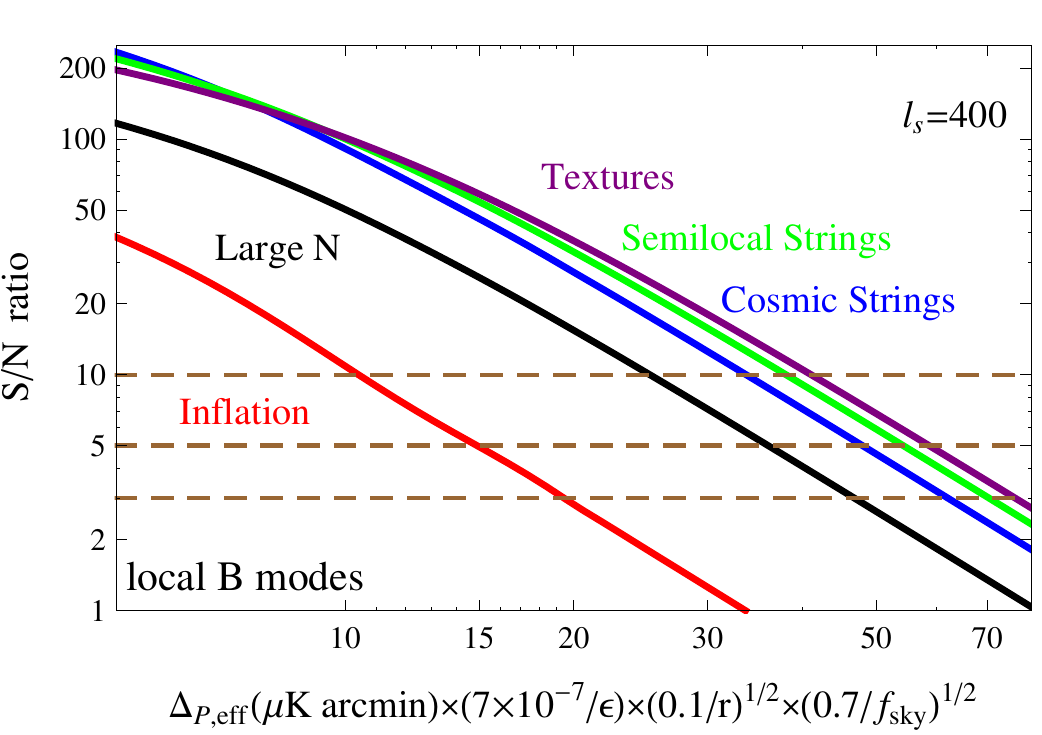}
}
\caption{The signal-to-noise ratio as a function of the normalized polarization sensitivity,
for inflation, cosmic strings, textures and the large-$N$ limit of the non-linear sigma-model.
}
\label{fig:SNR1}
\end{figure}

In a recent paper~\cite{GBDFFK} we analyzed the possibility of disentangling the different contributions 
to the B-mode polarization coming from defects versus that from inflation. The main difficulty, for both
defects and tensor modes from inflation, is that the B-mode power spectrum is ``contaminated" by the
effect of lensing from the intervening matter distribution on the dominant E-mode contribution on
similar angular scales. Using the temperature power spectrum to determine the underlying matter
perturbation from evolved large scale structures responsible for CMB lensing, it is possible to engineer
an iterative scheme to clean the primordial B-modes from lensed E-modes~\cite{SeljakHirata}. This procedure
leaves a significantly smaller polarization noise background $\Delta_{P,{\rm eff}}$ which allows one to 
detect the GW background at high confidence level (3-$\sigma$) if the scale of inflation or that of
symmetry breaking is high enough. What we realized is that the usual E- and B-modes used for
computing the angular power spectra are complicated non-local functions of the Stokes parameters,
involving both partial differentiation and inverse laplacian integration. Such a non-local function
requires knowledge of the global polarization on scales as large as the horizon, where the B-mode
angular correlation function is negligible and thus prone to large systematic errors. In contrast, the
so-called ``local" \~E- and \~B-modes~\cite{DurrerBook,BZ} can be constructed directly from the Stokes 
parameters and do not involve any non-local inversion. A direct consequence in this change of
variables is that the angular power spectrum of local \~B-modes has a extra factor 
$n_\ell = (\ell+2)!/(\ell-2)! \sim \ell^4$, which boosts the high-$\ell$ peak in the defects' power spectra.

\begin{figure}
\centerline{
\includegraphics[width=13.0 cm,height=10.0 cm]{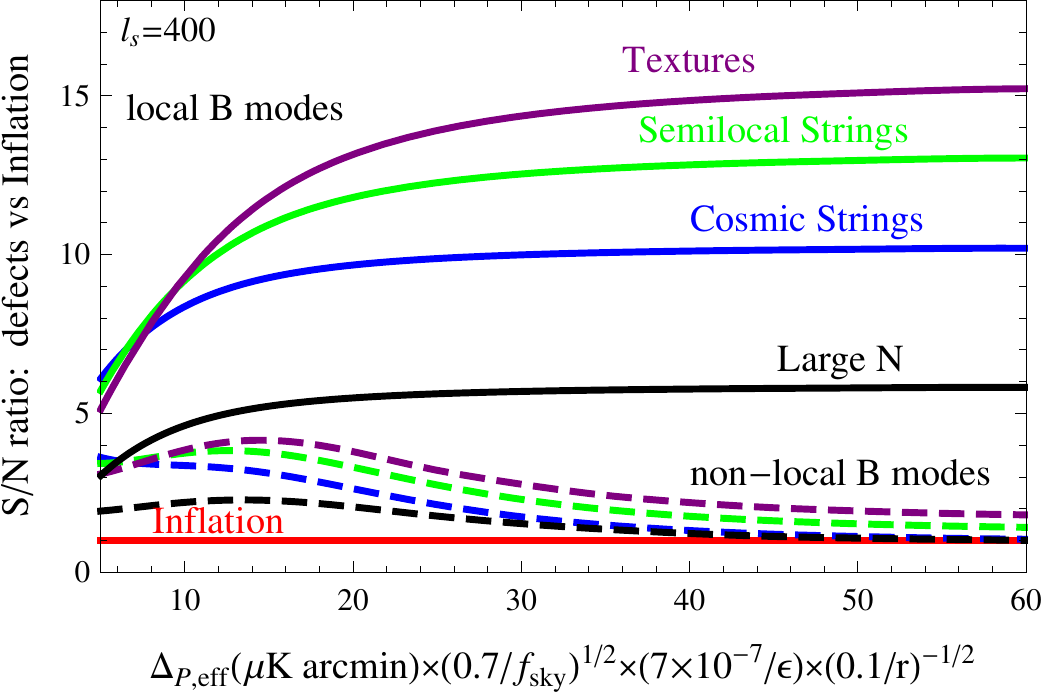}
}
\caption{The relative signal-to-noise ratio for defect models versus inflation for local (continuous lines)
and non-local (dashed lines) B-modes.}
\label{fig:SNR}
\end{figure}

When compared with the angular correlation function of inflation, it gives a significant advantage to the
defects' prospects for detection in future CMB experiments, see Fig.~\ref{fig:SNR}. In Table~1 we give the expected
bounds at 3-$\sigma$ on the tensor amplitude for inflation and various defect models, while in Table~2
we give the corresponding bounds on the scale of inflation, and on the symmetry breaking scale for defects, 
according to 
$$M_{\rm inf} = 1.63\times10^{16}\ {\rm GeV}\left(\frac{r}{0.1}\right)^{1/4}\,, \hspace{1cm}
v=1.02\times10^{16}\ {\rm GeV}\left(\frac{\epsilon}{7\times10^{-7}}\right)^{1/2}\,.$$ 
This gives the fundamental connection between the
CMB observations and the particle physics models at ultra high energy, much beyond any experiment will
ever be able to reach.

\begin{table}
\caption{ {The limiting amplitude for inflation (r=T/S) and various defects ($\epsilon=Gv^2$), 
at 3-$\sigma$ in the range $\theta\in[0,1^o]$, for Planck ($\Delta_{P,{\rm eff}}=
11.2\,\mu$K$\cdot$arcmin), COrE-like exp. ($\Delta_{P,{\rm eff}}=0.7\,\mu$K$\cdot$arcmin) 
and a dedicated CMB experiment with $\Delta_{P,{\rm eff}}=0.01\,\mu$K$\cdot$arcmin. We 
have assumed $f_{\rm sky}=0.7$ for all CMB experiments. } }
\label{tab:3sigma}
\vspace{0.4cm}
\begin{center}
\begin{tabular}{|c|ccccc|}
\hline
$S/N=3$ & Inflation & Strings & Semilocal & Textures & Large-N \\
\hline
\hline
Planck & $0.03$ & $1.2\times10^{-7}$ & $1.1\times10^{-7}$ & $1.0\times10^{-7}$ & $1.6\times10^{-7}$  \\
\hline
B-pol & $10^{-4}$ & $7.7\times10^{-9}$ & $6.9\times10^{-9}$ & $6.3\times10^{-9}$ & $1.0\times10^{-8}$  \\
\hline
$\tilde B$ exp & $10^{-7}$ & $1.1\times10^{-10}$ & $1.0\times10^{-10}$ & $0.9\times10^{-10}$ & $1.4\times10^{-10}$  \\
\hline
\end{tabular}
\end{center}
\end{table}

\begin{table}
\caption{ {The bounds on the scale of inflation and the symmetry breaking scale in units of GeV, 
at 3-$\sigma$ in the range $\theta\in[0,1^o]$, for Planck ($\Delta_{P,{\rm eff}}=
11.2\,\mu$K$\cdot$arcmin), COrE-like exp. ($\Delta_{P,{\rm eff}}=0.7\,\mu$K$\cdot$arcmin) 
and a dedicated CMB experiment with $\Delta_{P,{\rm eff}}=0.01\,\mu$K$\cdot$arcmin. We 
have assumed $f_{\rm sky}=0.7$ for all CMB experiments. } }
\label{tab:inflation}
\vspace{0.4cm}
\begin{center}
\begin{tabular}{|c|ccccc|}
\hline
$S/N=3$ & Inflation & Strings & Semilocal & Textures & Large-N \\
\hline
\hline
Planck & $1.2\times10^{16}$ & $4.2\times10^{15}$ & $4.0\times10^{15}$ & $3.9\times10^{15}$ & $4.9\times10^{15}$  \\
\hline
B-pol & $2.9\times10^{15}$ & $1.1\times10^{15}$ & $1.0\times10^{15}$ & $9.7\times10^{14}$ & $1.2\times10^{15}$  \\
\hline
$\tilde B$ exp & $5.2\times10^{14}$ & $1.3\times10^{14}$ & $1.2\times10^{14}$ & $1.1\times10^{14}$ & $1.4\times10^{14}$  \\
\hline
\end{tabular}
\end{center}
\end{table}

\section*{Acknowledgments} I thank Ruth Durrer, Jeff Dufaux, Elisa Fenu, Daniel G. Figueroa and Martin Kunz, 
for a very enjoyable collaboration. I specially thank the Institute de Physique Th\'eorique de l'Universit\'e de 
Gen\`eve for their generous hospitality during my sabbatical in Geneva. 
This work is supported by the Spanish MICINN under Project No. AYA2009-13936-C06-06, 
the CAM project ``HEPHACOS'' Ref. S2009/ESP-1473, and by the EU FP6 Marie Curie Research and 
Training Network ``Universe Net'' Ref. MRTN-CT-2006-035863.

\end{document}